\documentclass[%
reprint,
 amsmath,amssymb,
 aps,
 pra,
]{revtex4-2}

\usepackage{appendix}

\usepackage{graphicx}
\usepackage{dcolumn}
\usepackage{bm}
\usepackage{comment}

\usepackage{xcolor}

\begin{document}


\title{Efficient Dicke state distribution in a network of lossy channels}

\author{Wojciech Roga$^{1}$}
 \email{wojciech.roga@keio.jp}

\author{Rikizo Ikuta$^{2,3}$}

\author{Tomoyuki Horikiri$^4$}

\author{Masahiro Takeoka$^{1}$}
 \email{takeoka@elec.keio.ac.jp}

\affiliation{%
$^{1}$Department of Electronics and Electrical Engineering, Keio University, Hiyoshi, Kohoku, Yokohama 223-8522, Japan\\
$^2$Graduate School of Engineering Science, Osaka University, Toyonaka, Osaka 560-8531, Japan\\
$^3$Center for Quantum Information and Quantum Biology, Osaka University, Osaka 560-8531, Japan\\
$^4$Department of Physics, Graduate School of Engineering Science, Yokohama National University, Tokiwadai, Hodogaya, Yokohama 240-8501, Japan\\
} 


\date{\today}

\begin{abstract}
We analyze the generation of entanglement in a multipartite optical network. We generalize the twin-field strategy to the multipartie case and show that our protocol has advantageous rate-loss scalings of distributing $W$ states and Dicke states over the star networks. We provide precise theoretical formulas and quantitative performance analyses. Also analysis of the same protocol using Gaussian states as resources, which is a typical setup in many experimental tests, is provided. We numerically study the feasibility of the protocol in realistic experimental conditions with imperfect detectors. We observe that the advantage depends on the chosen fidelity and is affected by the dark counts of the detector.
\end{abstract}


\maketitle

\section{Introduction}


Quantum repeaters have been investigated for a long time 
in the context of overcoming the effect of loss in transmitting optical channels. 
They are a basic building block for extending quantum key distribution (QKD) networks without trusted nodes \cite{Lo, Pirandola} and a crucial step toward building a quantum internet \cite{Kimble, Wehner}. 
Various protocols for quantum repeaters have been explored, especially in the context of bipartite transmission. In many of those proposed so far, quantum memory \cite{Scherer, Lvovsky} or large-scale entangled photonic states \cite{Munro, Azuma} are required at the repeater nodes. 
These approaches still face serious 
technical challenges.  
Meanwhile, a recently proposed QKD protocol, the twin-field QKD (TF-QKD), shows that a single-hop repeaterlike operation is possible with a very simple configuration at the intermediate node \cite{Lucamarini}, with only linear optics and photon detection. 
While the TF-QKD protocol uses weak coherent-state signals, the corresponding entanglement-based protocol \cite{Curty} 
can also act as a single-hop repeater for entanglement distribution. 
This idea has been extended to a multiparty QKD scenario known as the conference key agreement \cite{Graselli}.
In \cite{Graselli}, the authors used $W$ states \cite{Dur} as a resource for QKD for the first time and showed that their protocol can achieve higher conference key generation rates than that of the direct transmission protocol. 

In this work, we further investigate in detail the single-hop repeater protocol for multipartite entangled states. 
We study the efficiency and quality of protocols generating $W$ and Dicke states that are characterized by advantageous scaling with respect to some nonrepeater benchmarks. 
We give the general formulas for an ideal protocol. 
In addition, we analyze realistic experimental conditions with Gaussian states generated from spontaneous parametric down-conversion sources.
Our work goes beyond the area covered by \cite{Graselli} in general characterizations of Dicke states, different nonrepeater benchmarks, and providing analysis of the protocols with Gaussian states and realistic experimental conditions, which is natural for existing experimental setups.

This paper is organized as follows. In Sec.~II, an overview of our idea and protocol is given. Section III describes the detailed theoretical formulation of the protocol. In Sec.~IV, we perform a numerical simulation of the protocol with practical Gaussian-state inputs. Section V concludes the paper. 



\section{Overview of the protocol}

Let us start by defining the states we want to generate in a network. Dicke states are $K$-photon pure states with at most one photon per mode that are invariant with respect to permutations of modes. We can also consider the generalized version, which can be brought to the standard Dicke states by local changes in phases,
\begin{equation}
|D_{(N,K)}\rangle = \frac{1}{\sqrt{C^N_K}}\sum_{k=1}^{C^N_K}e^{i\phi_k}|f^{(N,K)}_k\rangle.
\end{equation} 
Here, $C^x_y$ is the number of combinations of $y$ elements in the set of $x$ elements and $|f^{(N,K)}_k\rangle$ is the state with $N$ modes and $K$ photons with at most one photon per mode. The summation is over $|f^{(N,K)}_k\rangle$ numbered in a chosen order by index $k$. As an example we consider the Dicke state with $D_{(4,2)}$,
\begin{eqnarray}
|D_{(4,2)}\rangle & = & \frac{1}{\sqrt{6}}(|1100\rangle+|1010\rangle+|1001\rangle \nonumber\\ 
&& +|0110\rangle+|0101\rangle+|0011\rangle).
\end{eqnarray}
By setting $K=1$ we define generalized $W$ states,
\begin{equation}
|W_N\rangle = \frac{1}{\sqrt{N}}\sum_{k=1}^Ne^{i\phi_k}|f^{(N,1)}_k\rangle.
\end{equation} 
For example,
\begin{equation}
|W_{4}\rangle = \frac{1}{\sqrt{4}}(|1000\rangle+|0100\rangle+|0010\rangle+|0001\rangle),
\end{equation}
which we reduce to the single photon maximally entangled two-mode states when $N=2$, i.e., $|W_{2}\rangle=|\Psi\rangle$.
Hereafter, we omit the phase factor $e^{i\phi_k}$ since it is easily controlled by the local phase rotations. 
\\

\begin{figure}
    \centering
    \includegraphics[scale = .45]{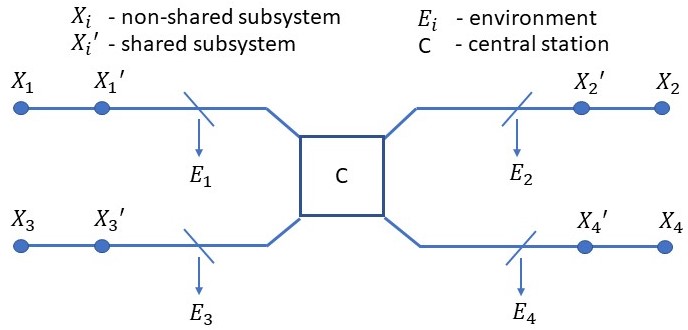}
    \caption{A network with $N=4$ parties connected to a central station $C$ by lossy channels, each of length $L$. The parties and the central station collaborate to generate Dicke states such that the generation rate scales favorably with the transmittance of the channel. $E_i$ denote the environments the photons can be lost to.}
    \label{fig:pic2}
\end{figure}

\begin{figure}
    \centering
    \includegraphics[scale = .45]{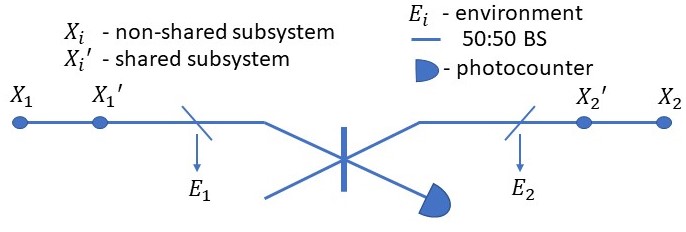}
    \caption{Generating one-photon entangled states in a network with two parties and a central station. The parties prepare entangled states similar to vacuum in their local sources and send one part of each to the central station that removes the information about the origin of potential photons. The detector's click denotes the creation of an entangled state between the parties. Success probability scales like the first power of the power transmittance of one channel that links one party with the central station. $E_i$ denote the environments the photons can be lost to.}
    \label{fig:pic}
\end{figure}

We consider a star network in which users and a central station are connected by identical lossy channels with power transmittance $T$ and introduce a protocol for entangled state generation that improves the success rate with respect to other protocols with a similar setup. The situation is schematically shown in Fig.~\ref{fig:pic2} for the case with four nodes. In our protocol, the nodes locally produce states of the form
\begin{equation}
|\psi\rangle_{X_iX_i'} = a|00\rangle_{X_iX_i'}+b|11\rangle_{X_iX_i'},
\label{initial}
\end{equation}
where $|i\rangle$ denotes an $i$-photon Fock state and the subsystems $X_i'$ are called shared subsystems and will be jointly measured in the later stage of the protocol. Here, we assume that $a$ and $b$ are real and positive; moreover, we assume that $b$ is small. The small value of $b$ is our choice. It is motivated by the fact that if the experiment signals success, the generated states have high fidelity with the target states, as is explained in detail later.
Then shared subsystems are sent to the middle station, pass through a mode-mixing setup that removes the information about where a photon comes from, and are measured (see Fig.~\ref{fig:pic}). When the appropriate measurement outcome appears, the nonshared system gets close to the Dicke state (detailed definitions will be given later) with $N$ modes and $K$ photons, which reduces to $W$ states when $K=1$ and to a maximally entangled bipartite state when $N=2$ and $K=1$. In particular the protocol allows for generating a maximally entangled state $ \frac{1}{\sqrt{2}}(|01\rangle_{X_1X_2}+|10\rangle_{X_1X_2})$ with the success probability that scales like $T$. This two-party scenario is an immediate consequence of the entanglement version of the TF-QKD protocol \cite{Lucamarini,Curty}. 
We will extend this to the multipartite scenario with $N>2$. 

Justification for our protocol comes from the following comparison to alternative methods for generating entangled states in the system. 
Let us first look at the case with $N=2$ and $K=1$, where two distant parties intend to share Bell states. 


 The trivial strategy is that one party generates two-photon Bell states, e.g. by polarization encoding, and then sends one part to the second party through the channel directly linking them. The process is characterized by the probability of success that a photon is not lost. This is given by the power transmittance. For example, for an optical fiber channel, it is given by, 
$$
T = 10^{-\gamma d/10},
$$
where $\gamma$ is the fiber loss coefficient typically given in decibels per kilometer. For the considered link of length $2d$ the probability that the entangled photon passes the channel is $T^2$.
This rate-loss scaling is, in fact, a fundamental limit of the distillable entanglement over a point-to-point lossy channel. In other words, one cannot beat this scaling by any input quantum signal with the assistance of any local operations and classical communication (LOCC) \cite{Takeoka,Pirandola17}.

Another strategy is to locate a central station in the middle of the channel, which generates entangled photon pairs, such as polarization entangled photons, and send them to each node. This is an experimentally established approach for entanglement distribution. Since each photon travels through a channel with loss $T$, the success probability of distributing them to two nodes is again $T^2$. 
This setting is out of the scope of point-to-point channel-capacity theorems \cite{Takeoka,Pirandola17}, and nontrivial tight bounds for its capacity are not known yet. 
In this paper, therefore, we use the above strategy of directly sending entanglement from the central station to benchmark our repeaterlike protocol and call it the direct-transmission protocol. 
This strategy can be easily extended to the multipartite-entanglement scenario. 
The central station generates $N$-partite entangled states with $N$ photons, e.g., by polarization encoding \cite{Pan,Eibl} and sends each photon to each user via a channel with loss $T$. 
Then the success probability of the distribution is given by $T^N$.  

\begin{figure}
    \centering
    \includegraphics[scale = .5]{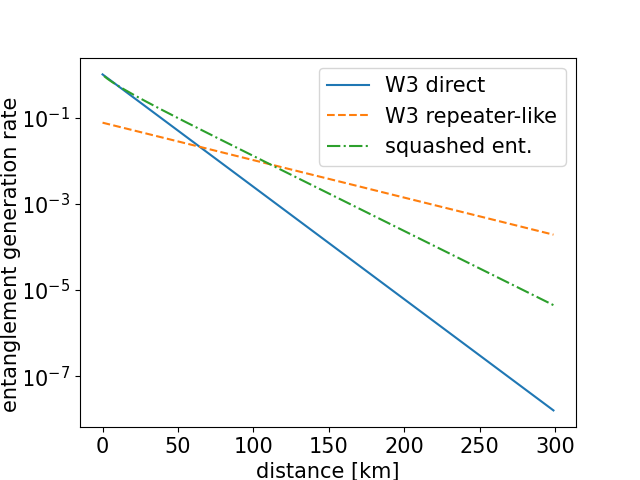}
    \caption{Comparison of our repeaterlike protocol, the direct transmission, and the squashed entanglement bound for the distribution of the tripartite $W$ state defined in (\ref{eq:w3}) in the star network. The generation rate is the probability of success of the experiment.
    The smaller rate of the repeaterlike scenario for short distances is implied by the low probability of generating one photon in the circuit, as $b$ is assumed to be small. However, the advantage of this method for larger distances comes from the fact that only one photon has to survive the loss.}
    \label{fig:squashed}
\end{figure}

Finally, let us also consider the rate of distributing three-party $W$ states consisting of one and zero photons: 
\begin{equation}
\label{eq:w3}
|W_{3}\rangle = \frac{1}{\sqrt{3}}(|100\rangle+|010\rangle+|001\rangle) .
\end{equation}
When it is distributed from the central station to distant parties, the state turns out to be a mixture of $|W_3\rangle$ and vacuum.
This means that one needs some distillation process to extract $|W_3\rangle$.  
The upper bound on the distillable rate of the $W$ state is, for example, given by the multipartite version of the squashed entanglement \cite{Yang}. 
In the Appendix A, we derive the upper bound of the squashed entanglement. In Fig.~\ref{fig:squashed}, we show plots comparing our method, the direct transmission of three-photon $W$ states, and the squashed entanglement-based bound for the direct transmission of $|W_3\rangle$ through the star network linked by a fiber with a loss of 0.2 dB/km (a typical loss parameter in the telecom band). The generation rate is the probability of success which determines the fraction of the experimental repetitions in which successful generation is signalized. The comparison shows that the method we propose provides advantageous scaling with respect to the others. (Note that the squashed entanglement bound shown here is not optimal and thus is not necessarily tight. It is an interesting open problem whether it can be further tightened.) 
The examples shown in this section motivate us to study our repeaterlike protocol in detail. 
In the following sections, we derive precise theoretical formulas and perform quantitative analyses for both ideal and realistic conditions.

\section{Generating generalized Dicke states}

\subsection{Description of the setup}

Suppose $N$ parties are connected to a central station with lossy optical channels, each with power transmittance $T$, (see Fig. \ref{fig:pic2}). We assume that each party $i$ can produce locally entangled photonic states (\ref{initial}) on their local subsystems $X_i$ and $X_i'$
and that the shared subsystems are sent to the central station. If $b$ is sufficiently small, there is mostly vacuum or one photon in the joint shared system. Let us assume that the central station $C$ is equipped with a circuit that completely removes the information about which mode the photon comes from.

\begin{figure}
    \centering
    \includegraphics[scale = .7]{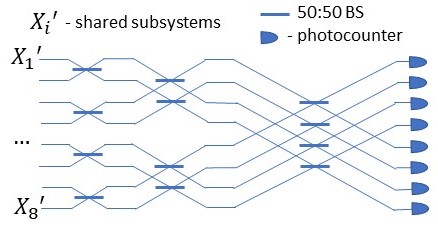}
    \caption{Scheme of the eight-mode transform that uniformly mixes a single-photon state. The short, thick horizontal lines denote 50:50 beam splitters. The scheme realizes the Hadamard transform on the modes' photon creation operators. }
    \label{fig:mixer} 
\end{figure}

For that purpose for $N = 2^n$ we propose a circuit of 50:50 beam splitters given as follows (see Fig.~\ref{fig:mixer}) \cite{Lougovski}. First, we group modes in pairs. We apply a beam splitter in each pair. Next, we group previous groups in pairs and apply a beam splitter to the corresponding outputs. We repeat this procedure  $n$ times. Formally, we can write this transform as a transformation of creation operators $(a_1^\dagger,...,a_N^\dagger)^T$ corresponding to modes as follows:
\begin{equation}
U = (u\otimes{\mathbf 1}_{n-1})({\mathbf 1}_{1}\otimes u\otimes{\mathbf 1}_{n-2}) ... ({\mathbf 1}_{n-1}\otimes u) = u^{\otimes n},
\end{equation}
where  ${\mathbf 1}$ is the two-dimensional identity matrix, ${\mathbf 1}_{i} = {\mathbf 1}^{\otimes i}$, and
\begin{equation}
u = \frac{1}{\sqrt{2}}\begin{pmatrix} 1&1\\-1&1\end{pmatrix}
\end{equation}
represents the beamsplitter. Notice that $U$ is the so-called Hadamard matrix, which realizes a real transform similar to the Fourier transform.
The circuit from Fig.~\ref{fig:mixer} was used in \cite{Lougovski} to generate $W$ states locally.

\subsection{Generation of $W$ states with ideal channels}

To explain the main idea of the protocol in simple terms, in this section, we consider only the generation of $W$ states with the setup without losses. The generation of Dicke states is a straightforward task and will be done in the next section together with a more technical analysis of the influence of losses. 

In the ideal case, i.e., with no losses in the channels, $T=1$, if parties send their shared subsystems to the station and only one photon is detected after the transform, the global state of the nonshared systems is projected onto the generalized $W$ state.
A feed-forward local compensation depending on which detector clicks can be performed to bring the generalized $W$ state to a chosen fixed-phase $W$ state. 
Indeed, the part of the state $|\psi\rangle_{X_1X'_1}\otimes ... \otimes|\psi\rangle_{X_NX'_N} $ before the transform with just one photon in the shared system is
\begin{equation}
\sum_{k=1}^N a^{N-1}b|f^{(N,1)}_k\rangle_{X}|f^{(N,1)}_k\rangle_{X'},
\end{equation}
where we simplified the notation by putting $X$ instead of $X_1,...,X_N$ and doing the same for the shared systems.
Each term of the shared system $X'$ is transformed into up-to-phase uniform superposition of single-photon states in all modes. So for each vector of the nonshared part $X$ we have exactly one term associated with a click of a chosen detector in $X'$. Hence, if this detector clicks, the nonshared part collapses to the generalized $W$ state. We can immediately give the probability of the click of a chosen detector, which is 
\begin{equation}
p_s = (a^{N-1}b)^2.
\end{equation} 
Note that because we keep $b$ 
small, the probability of multiphoton detection is small. However,
if more than one photon is detected, this method can be used to generate a Dicke state in the nonshared system. Generalization of the lossless protocol for Dicke-state generation is almost straightforward, so we will not discuss it separately. Instead, in the following section we will analyze the general scheme of Dicke-state generation with lossy channels.

\subsection{Dicke-state generation with lossy channels}

Let us consider in full generality generation of Dicke states $D_{(N,M)}$ in the setup with losses. Here, we deal with $M$ photons surviving the losses, going through the transform of the central station, and being detected by a configuration of detectors indexed by $s$. For instance, for $N=4$ modes and $M=2$ photons, $s$ can take the form $s=(0,1,1,0)$, which denotes that the second and third detectors click; $s=(2,0,0,0)$, which means that the first detector records two photons; and so on.

To deal with the problem systematically we consider parts of the state $|\psi\rangle_{X_1X'_1}\otimes ... \otimes|\psi\rangle_{X_NX'_N} $ associated with $K$ photons, where $K\geq M$. We have $C^N_K$ of these terms, all with the same amplitudes, $a^{N-K}b^K$, namely,
\begin{equation}
\sum_{k=1}^{C^N_K} a^{N-K}b^K|f^{(N,K)}_k\rangle_{X}|f^{(N,K)}_k\rangle_{X'}.
\end{equation}
We are interested in the part with $M$ photons left in the shared system and $K-M$ photons lost to the environment $E = E_1,...,E_N$, which is
\begin{eqnarray}
\sum_{k=1}^{C^N_K} a^{N-K}b^K(\sqrt{T})^M(\sqrt{1-T})^{K-M}|f^{(N,K)}_k\rangle_{X}\nonumber\\
\otimes\sum_{m=1}^{C^K_M}|g^{(M,K,k)}_m\rangle_{X'}
|g^{(K-M,K,k)}_m\rangle_{E},
\end{eqnarray}
where $|g^{(x,K,k)}_m\rangle$ denotes a vector with $x$ photons distributed in $K$ modes selected by index $k$, one photon per mode. Different vectors are numbered by index $m$, and the subscript $E$ refers to modes of the environment. The probability of this part of the global state gives us the probability of $M$ photons reaching the central station
\begin{equation}
p' = \sum_{K=M}^{N}\left[a^{N-K}b^K(\sqrt{T})^M(\sqrt{1-T})^{K-M}\right]^2C^N_KC^K_M
\end{equation}
for $T\leq 1$. 
The probability that after the transform $U$ at the central station a chosen configuration $s$ of detectors clicks is
\begin{eqnarray}
p_s & = & \sum_{K=M}^{N}\left(a^{N-K}b^K(\sqrt{T})^M(\sqrt{1-T})^{K-M}\right)^2\nonumber\\
&&\times\sum_{k=1}^{C^N_K}\sum_{m=1}^{C^K_M}|\langle g^{(M,K,k)}_m|U|s\rangle|^2,
\label{efficone}
\end{eqnarray}
where $\langle g^{(M,K,k)}_m|U|s\rangle$ is a permanent of a matrix built of elements from intersections of rows and columns of $U$ indicated by nonzero entries of $g^{(M,K,k)}_m$ and $s$, respectively. When $s$ contains multiple photons in a given mode, the appropriate column should be repeated in order to calculate the permanent.
In general the permanent given by $\langle g^{(M,K,k)}_m|U|s\rangle$ can take different values depending on $s$ and $g^{(M,K,k)}_m$; however, it is a constant when $M$ photons are detected in a single detector. Then $|\langle g^{(M,K,k)}_m|U|s\rangle|=\sqrt{M!}/\sqrt{N}^M$. 
Assuming this and expanding $p_s$ close to $T=0$ up to $O(T^{2M})$, we find a concise formula for the efficiency of this process,
\begin{equation}
p_s=T^{M}b^{2M}\frac{C^N_MM!}{N^M}.
\label{efficshort}
\end{equation}
A detailed derivation of this formula is given the Appendix B. 

The nonshared system is then projected onto a state. The fidelity of this state with a generalized Dicke state is
\begin{equation}
F = \frac{1}{p_s}a^{2(N-M)}b^{2M}T^{M}\left|\sum_k^{C^N_M}\frac{1}{\sqrt{C^N_M}}\left|\langle f^{(N,M)}_k|U|s\rangle\right|\right|^2.
\label{fidelone}
\end{equation} 
For small $T$, taking into account (\ref{efficshort}), the fidelity is
\begin{equation}
F = (a^2)^{N-M}.
\end{equation}
Having this general formula, we can express the probability that one of the detectors clicks for $M=1$ in the approximation up to $O(T^2)$ as a function of fixed fidelity:
\begin{equation}
p = N(1-F^{\frac{1}{N-1}})T.
\label{Fconst}
\end{equation}
Here, the factor $N$ comes from the fact that there are $N$ detectors with the same probability of a click given by (\ref{efficshort}). As a click of different detectors signalizes a $W$ state with possibly different phases, the above rate is achieved if a proper feed-forward correction is applied as follows. Let us assume that index $k$ indicates the position of one photon and index $s$ denotes the detector that fired, i.e., $s=(1,0,0)$ if the first detector fired and so on. Assume also that we want to bring the  generalized $W$ state to the $W$ state in which all phases are positive. Then, if the sign of the matrix element $\langle s|U|f^{(N,1)}_k\rangle$ is negative, the user $k$ should apply a phase shifter which changes the phase of one photon state by $\pi$.   

\begin{figure}
    \centering
    \includegraphics[scale = .5]{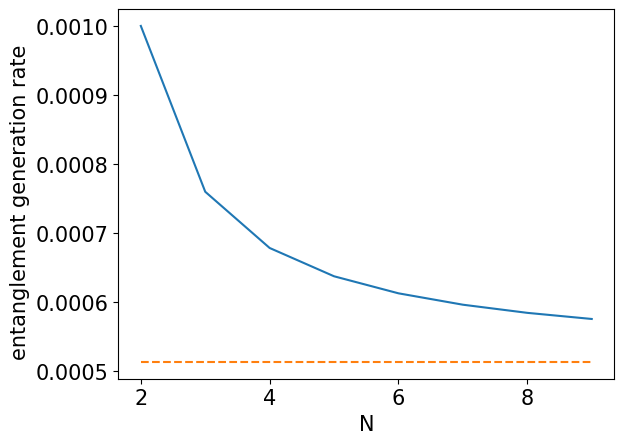}
    \caption{Generation rate of an $N$-partite state $W$ state with fidelity $F$ in the setup in which the parties are connected to the central station by a lossy channel with transmittance $t=\sqrt{T}$. The dashed line shows the asymptote. }
    \label{fig:probverN} 
\end{figure}

For $F$ close to 1 and $T$ close to 0 the probability is a monotonically decreasing function that asymptotically tends to a constant value, $\lim_{N\rightarrow\infty}N(1-F^{\frac{1}{N-1}})T=T\ln (1/F)$. 
For fixed values of the fidelity and transmittance of the channels the behavior of (\ref{Fconst}) as a function of $N$ and the asymptote are shown in Fig. \ref{fig:probverN}.

\begin{figure}
    \centering
    \includegraphics[scale = .5]{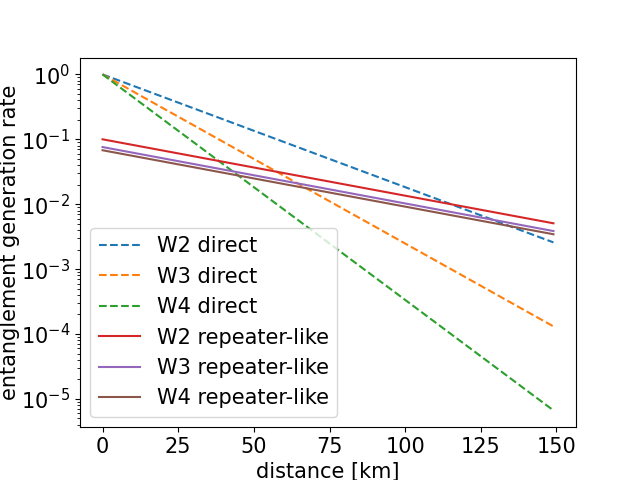}
    \caption{Comparison of the generation rate of $W_i$ states for $i=2,3,4$ with the direct method and using the repeaterlike method. Dashed and solid lines corresponding to $W_2$ have the highest generation rate, while the lines corresponding to $W_4$ have the lowest. For the repeaterlike scenario we assume fidelity $F=0.95$. We assume the loss rate is $0.2\ {\rm dB/km}$. In the direct method we consider that a central station sends parts of the $i$-photon polarization $W$ state to each node.}
    \label{fig:Wcomparison} 
\end{figure}

In Fig. \ref{fig:Wcomparison} we show a comparison of the generation rate (success probability) of $W_i$ states for $i=2,3,4$ between the direct method and using the strategy with the central detection. In the direct method we assume that a central station sends parts of the $i$-photon polarization $W$ state to each node, as discussed in the Introduction. We conclude that the scaling of the repeater-like method for long distances is favorable. Over short distances the direct method has a larger probability of success because the photon loss is smaller than the probability of having at least one photon detected. However, for larger distances in the direct method a photon must survive two times the distance of each of the photons that can be generated in the repeaterlike method. So in the case of the latter protocol, there are two contributions to the probability of success: generation of the photon by the source and the chances of one of the photons surviving the path to the central station. The probability of one photon being generated by the source is the overhead cost of the method that influences its efficiency for small distances, but in terms of scaling it does not matter. The fact that the probability of generating a $W$ state depends on the probability that only one of the photons survives one path is reflected in the weak dependence of the generation rate on the number of users in the repeaterlike scheme, as shown in Fig. \ref{fig:Wcomparison}.

\begin{figure}
    \centering
    \includegraphics[scale = .5]{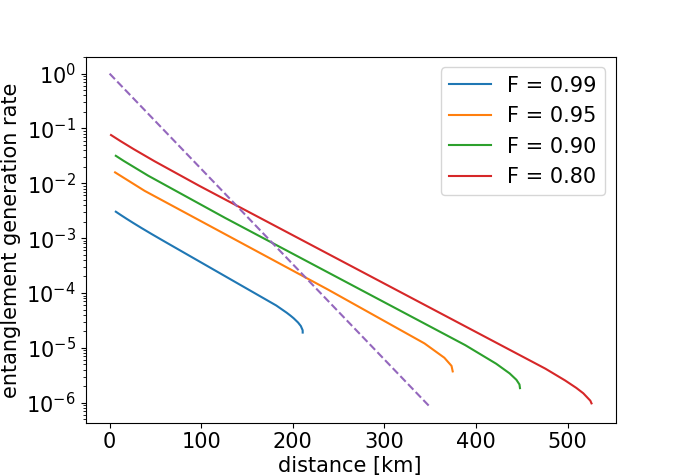}
    \caption{Comparison of the generation rate of different four-mode states using the direct method and the repeaterlike method. For the repeater-like method we assume fidelity $F=0.95$. We assume the loss rate is $0.2\ {\rm dB/km}$. In the direct method we consider that a central station sends parts of a four-photon polarization $W$ state to each node. In this plot lines corresponding to higher fidelity have a lower generation rate.}
    \label{fig:compar4} 
\end{figure}

In Fig. \ref{fig:compar4}  we compare generation rates (success probabilities) for four-mode $W$ and Dicke $D_{(4,2)}$ states using the direct method and our repeaterlike method. Again, we conclude that the repeaterlike method, although it has disadvantageous overhead cost related to the probability of generating one photon at the source, scales advantageously for longer distances, where the overhead does not matter.

\subsection{Generating two- and three-photon Dicke states}

Let us discuss separately the case with $N=4$ and $M=2$ when two different detectors click (which is the opposite of the previous situation in which the same detector recorded two photons). Direct calculations show that the probability of two different chosen detectors clicking is equal to
\begin{equation}
p_s = \frac{1}{2}a^4b^4T^2+a^2b^6T^2(1-T) +\frac{1}{2}b^8T^2(1-T)^2.
\end{equation}
If we expand this formula close to $T=0$, we get
\begin{equation}
p_s=\frac{1}{2}T^2b^4+O(T^3). 
\end{equation}
Substituting the first term in the formula for fidelity, we get
\begin{equation}
F = \frac{1}{3}a^4 \leq \frac{1}{3}.
\end{equation}
Hence, coincidence measurements with two different detectors do not project the nonshared system arbitrarily close to a Dicke state. We can understand this fact by observing that our transform with coincidence detection only partially removes the information about the input modes. Indeed, if photons came from modes 1 and 2 and the first detector clicks, then the second detector must click because of bunching photons at the first beamsplitter. However the situation changes if we observe two photons arriving at the same detector. The probability of this event for any detector is
\begin{eqnarray}
p_s &=& \frac{3}{4}a^4b^4T^2+\frac{3}{2}a^2b^6T^2(1-T) +\frac{3}{4}b^8T^2(1-T)^2\nonumber\\
&=&\frac{3}{4}T^2b^4+O(T^3).
\end{eqnarray}
That leads to fidelity between the state of the nonshared system and the Dicke state in the expansion up to $T^2$
\begin{equation}
F=a^4,
\end{equation}
which can be kept close to 1.

The situation with three photons in four modes (or even eight modes) is different because any three-photon detection projects  the nonshared system on a state arbitrarily close to a Dicke state.

\section{Generating $W$ states with input Gaussian states}

In this section we analyze a situation with two, three and four modes in which we want to generate a state close to the single-photon $W$ state with the appropriate number of modes in the nonshared system:
\begin{eqnarray}
|W_2\rangle &=& \frac{1}{\sqrt{2}}(|01\rangle+|10\rangle),\label{maxentst}\\
|W_3\rangle &=& \frac{1}{\sqrt{3}}(|001\rangle+|010\rangle+|100\rangle),\\
|W_4\rangle &=& \frac{1}{\sqrt{4}}(|0001\rangle+|0010\rangle+|0100\rangle+|1000\rangle),
\end{eqnarray}
respectivelym. We assume that instead of states (\ref{initial}),
parties prepare locally two-mode squeezed vacuum states
\begin{equation}
    |\psi\rangle_{X_iX_i'} = \sqrt{1-\lambda^2}\sum_{n=0}^\infty\lambda^n|nn\rangle_{X_iX_i'},
\end{equation}
where $\lambda = \tanh r$ and $r$ is the squeezing parameter. For small $r$ the state is mostly vacuum with a small addition of terms with nonzero photon numbers. As in the cases considered previously, the shared subsystems $X_i'$ go through lossy channels. In each channel the loss is modeled by a beam splitter with vacuum in the extra input mode. Then the shared subsystems $X_i'$ are mixed in the central station circuit that erases information about the origin of the photons. For the case with two parties, this is a 50:50 beam splitter as in Fig. \ref{fig:pic}. For three or four parties we use a circuit with four input and four output modes, which is a reduced version of the circuit in Fig. \ref{fig:mixer}. In the case of three users we set vacuum as the input of the fourth port. Finally, we observe a click of one of the detectors ($X_1'$) while other output modes are ignored. In what follows we assume detectors are not photon-number resolving; that is, the detector click refers to one or more photons, or the dark count. 
We assume that the probability of the dark count $p_{dc}=10^{-7}$ and the efficiency of the detector is $0.8$.

Under these conditions we calculate the probability of the click and the fidelity of the conditional state of the non-shared system collectively denoted as $X$ with a perfect $W_N$ state for $N=2,3,4$. The probability is obtained from the phenomenological formula including the dark counts, 
\begin{equation}
p = 1 - p_0  + p_{dc},    
\end{equation}
where $p_0$ is the probability that the detector does not click, and $p_{dc}$ is the dark-count probability. Here,
\begin{equation}
    p_0 = {\rm Tr}_{XX'E}\,\rho_{XX'E}|0\rangle\langle 0|_{X_1'},
\end{equation}
where $X$ denotes the subsystem that was not sent to the circuit and $X'$ is the subsystem sent to the circuit. (This circuit consists of $X_1'X_2'$ for the case with two parties and $X_1'...X_4'$ for the case with three or four parties. In the case of three parties the input to mode $X_4'$ is the vacuum.) 
Subsystem $E$ corresponds to the environment related to losses. Note that $\rho_{XX'E}$ is a Gaussian state and can be fully described by a covariance matrix. We can also calculate $p_0$ from known formulas in the Gaussian formalism.   


In the phenomenological model, we assume that when a dark count occurs, we do not get any information about the shared subsystem; therefore, we trace it out. Thus, after the detector reports the event, our nonshared system state takes the form
\begin{equation}
\sigma_{X}=\frac{1}{\mathcal{N}}\left[(p_{dc}+1)\rho_{X} -{\rm Tr}_{X'}(\rho_{XX'}|0\rangle\langle 0|_{X_1'})\right],
\end{equation}
where $\mathcal{N}=p_{dc}+(1-p_0)$ is the normalization. We calculate the fidelity between the state $\sigma_X$ and $|W_N\rangle$ by calculating density-matrix elements \cite{Kruse, xanadu2019} of $\rho_X$ and ${\rm Tr}_{X'}(\rho_{XX'}|0\rangle\langle 0|_{X_1'})$ corresponding to each single-photon term of $|W_N\rangle$. For example, for $N=3$ we need to calculate $\langle 001|\rho_3|010\rangle$ and all other terms with photons in different modes and to do the same for ${\rm Tr}_{X'}(\rho_{XX'}|0\rangle\langle 0|_{X_1'})$. We do that using the function ${\rm DENSITY\_MATRIX\_ELEMENTS}$ from the PYTHON library ${\rm THEWALRUS.QUANTUM}$ provided by Xanadu \cite{xanadu2019}.

Assuming transmission channels with a loss rate of $0.2 {\rm dB/km}$, we calculate the rate and fidelity for the generation of $W_N$, where $N=2,3,4$. The results are presented in the plots below.

\begin{figure}
    \centering
    \includegraphics[scale = .5]{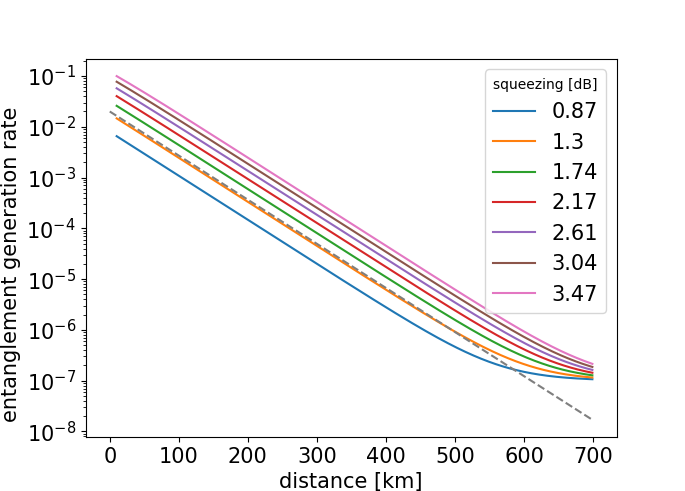}
    \caption{Probability of a single detector click (generating the $W_2$ state from Gaussian states) as a function of distance for a loss rate of $0.2\ {\rm dB/km}$ for different squeezings $(0.87, 1.3, 1.74, 2.17, 2.61, 3.04, 3.47 {\rm dB})$. The lowest line corresponds to the smallest squeezing. For comparison the dashed line shows formula (\ref{efficshort}) with a fidelity of $0.99$.}
    \label{fig:compar5} 
\end{figure}

\begin{figure}
    \centering
    \includegraphics[scale = .5]{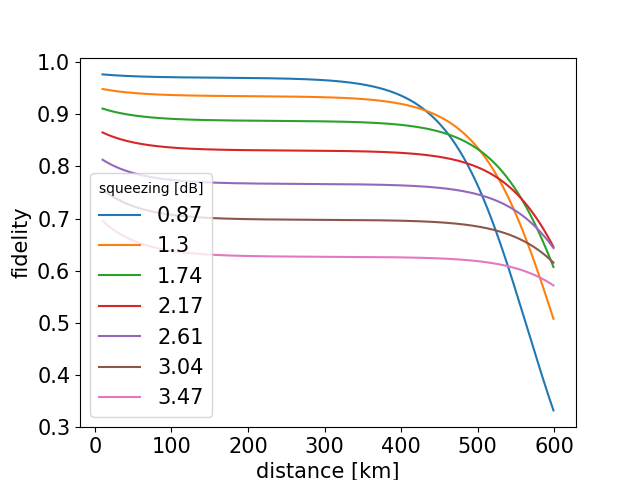}
    \caption{Fidelity of the $W_2$ state with the generated state when a single detector clicks as a function of distance for a loss rate of $0.2\ {\rm dB/km}$ for different squeezing parameters $(0.87, 1.3, 1.74, 2.17, 2.61, 3.04, 3.47 {\rm dB})$. The blue line (with the highest fidelity for small distances) corresponds to the smallest squeezing.}
    \label{fig:compar6} 
\end{figure}

Figure \ref{fig:compar5} compares the rate of generating $W_2$ versus distance for different squeezings. We follow the convention of experimental studies and express squeezing in decibels using
\begin{equation}
    10\log_{10}e^{2r} {\rm dB}
\end{equation}
where $e^{-2r}$ gives the amount of Einstein-Podolsky-Rosen correlation in the vacuum squeezed state \cite{Adesso}
\begin{equation}
    Y = \frac{1}{2} \left[{\rm Var}(x_1-x_2)+{\rm Var}(p_1+p_2)\right],
\end{equation}
and $(x_i, p_i)$ are quadrature operators. 
Figure \ref{fig:compar6} shows the fidelity versus distance. The rate decreases with distance, until the dark count dominates the rate of clicks. In that region we observe a strong decrease in fidelity. 


\begin{figure}
    \centering
    \includegraphics[scale = .5]{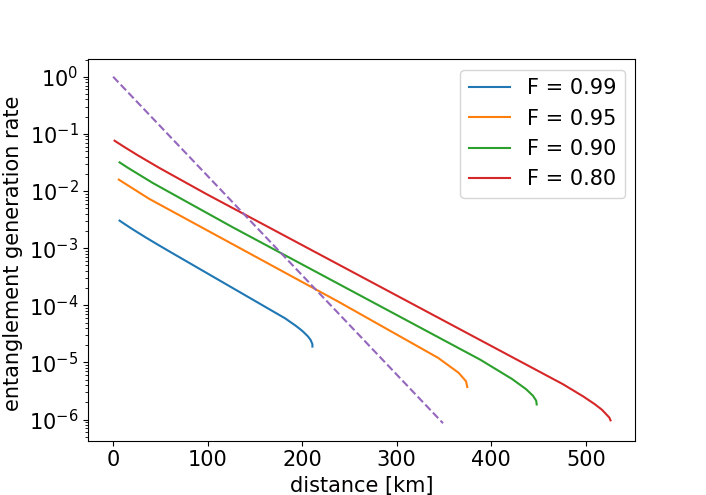}
    \caption{Generation rate of fixed-fidelity $W$ states from Gaussian states as a function of the distance. The dashed line shows the rate of generation of $W$ states using the direct method. The line corresponding to the lowest fidelity has the highest rate.}
    \label{fig:compar8} 
\end{figure}

\begin{figure}
    \centering
    \includegraphics[scale = .5]{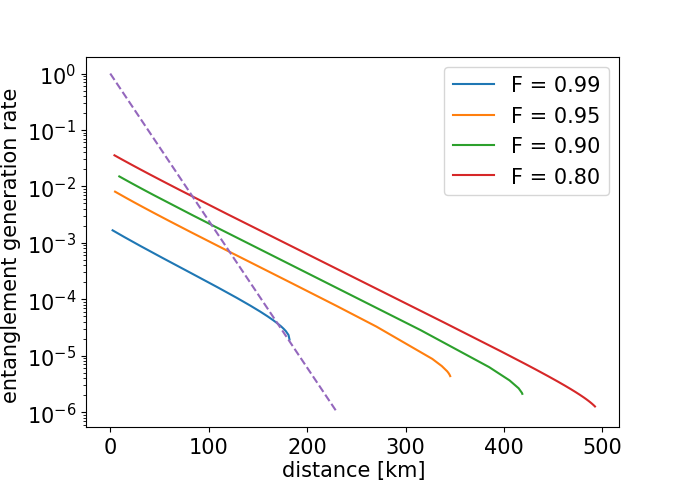}
    \caption{Rate of generation of $W_3$ with different fidelities. The dashed line denotes the direct method. The line corresponding to the lowest fidelity has the highest rate.}
    \label{fig:comp10} 
\end{figure}

\begin{figure}
    \centering
    \includegraphics[scale = .5]{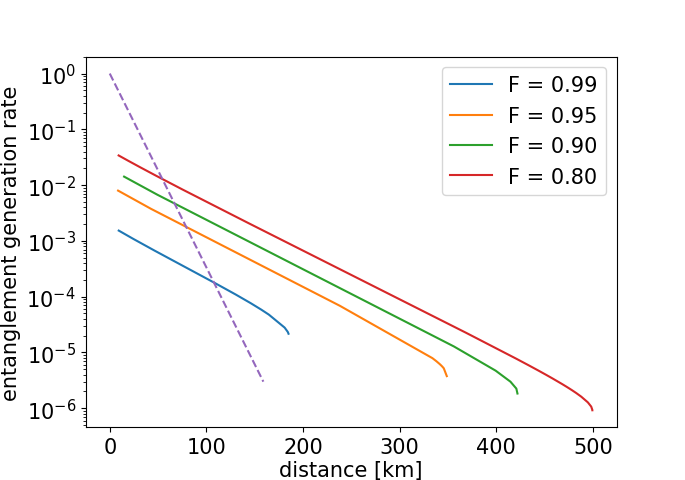}
    \caption{Rate of generation of $W_4$ with different fidelities. The dashed line denotes the direct method. The line corresponding to the lowest fidelity has the highest rate.}
    \label{fig:comp11} 
\end{figure}

In Figs.~\ref{fig:compar8}, \ref{fig:comp10}, and \ref{fig:comp11} we show the dependence of the rate on the distance for $W_2$, $W_3$, and $W_4$, respectively, for various values of fidelity. For comparison we also show the scaling of the rate for direct sharing of entanglement. We observe that the scaling of the method we investigate is better than that of the direct method. The highest range is achieved for the lower fidelity, of course. We observe that there is a fidelity level for which the direct method cannot be beaten for the quality of the detector that we assume. Namely, the quality of the detector is mainly the dark count, which is assumed to be equal to $10^{-7}$. For large distances the dark counts are more likely than the real signal to be affected by the loss; this is the reason why the efficiency sharply drops in Figs. 10-12. The value of this limiting fidelity is around 0.97 for $W_2$, 0.99 for $W_3$, and higher than 0.99 for $W_4$.
 
\begin{figure}
    \centering
    \includegraphics[scale = .5]{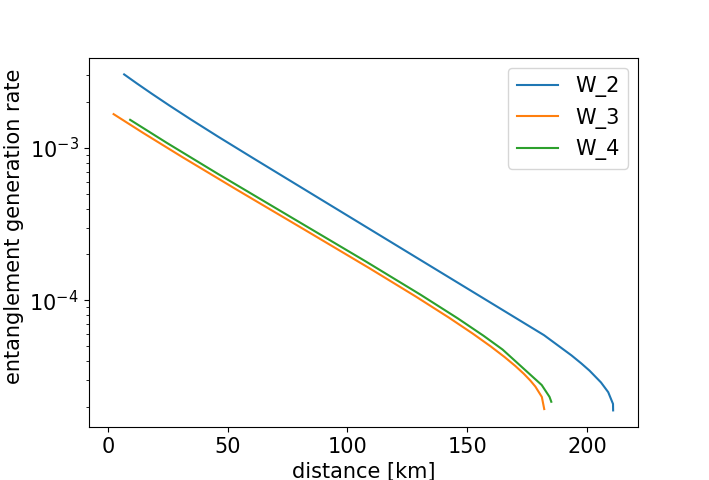}
    \caption{Generation rate for $W_N$ with fidelity $F=0.99$. The line corresponding to $W_2$ has the highest rate. The line for $W_3$ has the lowest.}
    \label{fig:comp12} 
\end{figure}
In Fig.~\ref{fig:comp12} we compare the rate at which we generate states $W_N$ with fixed fidelity $F=0.99$ for $N=2,3,4$. To achieve the given fidelity we apply different squeezings to the initial state. The generation rate of $W_2$ with fixed fidelity is the highest. In this case the circuit erasing the information about the paths of from which photons arrive is the simplest and consists of one beam splitter and two output modes. When we move from two to three or four parties, we use the circuit with four beam splitters and four detectors. We observe that the rate of generating a fidelity of $0.99$ for $W_3$ is lower than the rate for $W_4$. We think this is because the circuit with four modes we used is not optimal for $W_3$. That is, choosing an optimal circuit could improve the rate for $W_3$. Although studying this case would be interesting, it is out of the scope of our current work.

Let us also comment on the choice of the squeezing parameter. Note that by increasing squeezing we increase the rate of success (Fig. \ref{fig:compar5}), but the fidelity decreases (\ref{fig:compar6}). So a trade-off relation between the two quantities exists, and the choice of the squeezing depends on what is prioritized in each situation. If we fix the distance, we observe that there are two different values of the squeezing corresponding to a given fidelity. We should choose the one for which the rate is larger. In the given conditions the dependence of the fidelity or rate can be determined following the procedures described above. 

\section{Conclusion}


In this paper we analyzed the generation of Dicke states in a lossy network with an arbitrary number of parties equally distant from a central station. We showed that this method provides advantageous scaling of the rate in terms of the power transmittance with respect to the direct link. We derived general formulas for the fidelity and rate of this process. Also, we provided an analysis of the rate and fidelity of the protocol when the input state is a two-mode squeezed vacuum instead of entangled two-photon states.

Interesting open problems remain. We compared our protocol with the simple direct-transmission protocol in which the central station emits the target states, such as $W$ states, encoded in polarized photons and computed the rate at which all photons reach the users simultaneously. This corresponds to typical experiments of multipartite entangled photonic state generation and distribution performed in many experimental laboratories with post-selection. However, in theory, we can think of its generalization such that the central station can generate a more sophisticated state to be sent and then users can distill the target state via LOCC. This may give a more fundamental limit of the repeaterless multipartite-entanglement distribution in the star network. 
From an experimental point of view, to implement our protocol, one needs to lock the phases of all states since it uses the superposition of vacuum and one photon. This might be a technical challenge, but recent experimental progress is promising in regard to the solution \cite{Casper}.

\begin{acknowledgments}
R.I., T.H., and M.T. acknowledge the members of the Quantum Internet Task Force for comprehensive and interdisciplinary discussions on the quantum internet.
R.I., T.H., and M.T. are supported by JST Moonshot R\&D Grant No. ~JPMJMS226C. 
W.R. and M.T. are supported by JST CREST Grant No.~JPMJCR1772 and JST Moonshot R\&D Grant No.~JPMJMS2061.
R.I. is supported by JST Moonshot R\&D Grant No.~JPMJMS2066, 
and MEXT/JSPS KAKENHI JP20H01839 and JP21H04445. W.R. and
M.T. are also supported by JST COI-NEXT Grant No.
JPMJPF2221.
\end{acknowledgments}

\appendix
\section{Squashed entanglement bound for the direct transmission of tripartite $W$ states}

In this appendix we calculate the so-called squashed entanglement that provides an upper bound on the rate of distilling $W$ states shared by the direct transmission in the star network. More specifically, we consider the situation in which the central station of the star network repeatedly generates 
\begin{equation}
|W_{3}\rangle = \frac{1}{\sqrt{3}}(|100\rangle+|010\rangle+|001\rangle)
\end{equation}
and transmits them through identical lossy channels to three receivers. Here, $|i\rangle$ is an $i$-photon state, and we assume a loss rate of $0.2 {\rm dB/km}$. Each of the states is degraded at the receiver, but the receivers can distill some $W$ states from the larger number of received states by using local operations and classical communication (LOCC). 

To upper bound the distillation rate, we use the multipartite squashed entanglement \cite{Yang,Avis}. 
Following \cite{Avis,Seshadreesan}, we define it as
\begin{equation}
E_{sq} (A_1;...;A_m)_\rho=\frac{1}{2} \inf_{S_ {E\rightarrow E'}}I(A_1;...;A_m|E'),
\label{squashed}
\end{equation}
where $I(A_1;...;A_m|E')$ is the multipartite conditional mutual information, defined as
$$
I(A_1;...;A_m|E)= \sum_{i = 1}^m H(A_i | E) - H(A_1... A_m|E),
$$
and $H$ is the conditional von Neumann entropy,
$$
H(X|Y) =  H(XY) - H(Y).
$$
Here von Neumann entropy $H(X) = - {\rm Tr} X \log_2 X$. The optimization in (\ref{squashed}) goes over so-called squashing channels in the environment involved in the evolution of the open quantum system. 
The multipartite squashed entanglement is known to be an entanglement measure which is LOCC monotone, asymptotically continuous, and additive to tensor-product states. 
This is known as an upper bound for the distillation rates of Greenberger-Horne-Zeilinger states and multipartite keys \cite{Yang,Seshadreesan}.
With an appropriate normalization factor, it can also upper bound the distillation rate of $m$-partite $W$ state as  
\begin{equation}
R_W \le \frac{2}{m \, h_2(1/m)} E_{sq} ( A_1;...; A_m)_\rho,  
\end{equation}
where $h_2(x)$ is the Shannon binary entropy. 

By $\eta$ we denote the transmittance of each channel connecting the central station with the receiver. Then the one copy of the initial $W$ state becomes the following state of the system of receivers $S$ and the environment $E$ where the photon can escape: 
\begin{equation}
|\Psi'\rangle = \sqrt{\eta}|W\rangle_S|000\rangle_E+\sqrt{1-\eta}|000\rangle_S|W\rangle_E.    
\end{equation}
We do not consider here the optimal squashing channel (thus, we consider an upper bound of the squashed entanglement). For the channel we apply 50:50 beamsplitter mixed environment modes with additional environment space $E'$. The full state reads
\begin{eqnarray}
|\Psi''\rangle &=& \sqrt{\eta}|W\rangle_S|000\rangle_E|000\rangle_{E'}+\sqrt{\frac{1-\eta}{2}}|000\rangle_S|W\rangle_E|000\rangle_{E'}\nonumber\\
&+&\sqrt{\frac{1-\eta}{2}}|000\rangle_S|000\rangle_E|W\rangle_{E'}.    
\end{eqnarray}
Next, we trace $E'$ out, obtaining
\begin{eqnarray}
    \rho_{SE}&=& \eta|W\rangle\langle W|\otimes |0\rangle\langle 0|\nonumber\\
    &+&\sqrt{\frac{\eta(1-\eta)}{2}}(|W\rangle\langle0|\otimes|0\rangle\langle W|+|0\rangle\langle W|\otimes|W\rangle\langle 0|)\nonumber\\
    &+&\frac{1-\eta}{2}|0\rangle\langle 0|\otimes(|0\rangle\langle 0|+|W\rangle\langle W|),
\end{eqnarray}
where we use simplified notation in which $|0\rangle$ represents $|000\rangle$.
Here $S$ consists of $A_1$, $A_2$, and $A_3$, and the state above is symmetric with respect to exchanging the subsystems $A_i$. For each of the subsystems we have
\begin{eqnarray}
\rho_{A_iE} &=& \frac{\eta}{3}|1\rangle\langle 1|\otimes |0\rangle\langle 0|+\frac{3+\eta}{6}|0\rangle\langle 0|\otimes |0\rangle\langle 0|\nonumber\\
&+&\sqrt{\frac{\eta(1-\eta)}{6}}(|1\rangle\langle 0|\otimes |0\rangle\langle W|+|0\rangle\langle 1|\otimes |W\rangle\langle 0|)\nonumber\\
&+&\frac{1-\eta}{2}|0\rangle\langle 0|\otimes |W\rangle\langle W|,\\
\rho_S&=&\eta|W\rangle\langle W|+(1-\eta) |0\rangle\langle 0|,\\
\rho_{A_i}&=&\frac{\eta}{3}|1\rangle\langle 1|+ \frac{3-\eta}{3}|0\rangle\langle 0|,\\
\rho_E&=&\frac{1+\eta}{2}|0\rangle\langle 0|+ \frac{1-\eta}{2}|W\rangle\langle W|.
\end{eqnarray}
The nonzero eigenvalues of $\rho_{SE}$ are $\{(1-\eta)/2,(1+\eta)/2\}$, and the nonzero eigenvalues of $\rho_{A_iE}$ are  $\{(3-\eta)/6,(3+\eta)/6\}$. Hence the squashed entanglement upper bound for the $W$ state reads
\begin{equation}
    \frac{1}{3 \, h_2(1/3)}\left[3H(\rho_{A_iE})-2H(\rho_E)-H(\rho_{SE})\right].
\end{equation}
The squashed entanglement is shown in Fig. \ref{fig:squashed}. It is compared with the rate for the protocol with central detection discussed in this paper in which the parties send half of the bipartite entangled states $|\Phi\rangle$ to the central station and the detection of one photon signalizes $W$ states in the remaining system.

\section{Low-transmittance Dicke-states generation rate}

Expansion of (\ref{efficone}) close to $T=0$ that leads to formula (\ref{efficshort}). We assume that $b$ is real and $a = \sqrt{1-b^2}$:

\begin{eqnarray}
  p_s&=&\frac{M!}{N^M}\sum_{K=M}^N (a^{N-K}b^K)^2 T^M C^N_K C^K_M\nonumber\\
     &=&T^M  b^{2M}  \frac{C^N_M M!}{N^M}\sum_{K=M}^N a^{2(N-K)}b^{2(K-M)}  C^{N-M}_{K-M}\nonumber\\
     &=& T^M  b^{2M} \frac{C^N_M M!}{N^M}\sum_{K-M=0}^{N-M} a^{2(N-K)}b^{2(K-M)}  C^{N-M}_{K-M}\nonumber\\
     &=&T^M  b^{2M} \frac{C^N_M M!}{N^M} (a^2+b^2)^{N-M}\nonumber\\
     &=&T^M  b^{2M} \frac{C^N_M M!}{N^M}. \nonumber
\end{eqnarray}

\end{document}